\documentclass[aps,longbibliography,showpacs,twocolumn,superscriptaddress,amsmath,amssymb,verbatim]{revtex4-2}

\usepackage[thinlines]{easytable}
\usepackage{graphicx}
\usepackage{lmodern}
\usepackage{epstopdf}
\usepackage{dcolumn}
\usepackage{bm}
\usepackage{subfigure}
\usepackage{makecell}
\usepackage{amsmath} 
\usepackage[pdftex,colorlinks=true,citecolor=blue,linkcolor=blue,urlcolor=blue,bookmarks=true]{hyperref}
\usepackage{mathtools}
\usepackage{braket}

\usepackage{color}
\usepackage{textcomp}
\definecolor{red}{rgb}{1,0,0}

\definecolor{blue}{rgb}{0,0,1}

\definecolor{green}{rgb}{0,1,0}

\begin{document}
	\preprint{APS}

\title{Magnetism and field-induced effects in the $S$ = 5/2 honeycomb
	lattice antiferromagnet FeP$_{3}$SiO$_{11}$
 }

\author{J. Khatua}
\affiliation{Department of Physics, Indian Institute of Technology Madras, Chennai 600036, India}
\affiliation{Department of Physics, Sungkyunkwan University, Suwon 16419, Republic of Korea}
\author{M. Gomilšek}
\affiliation{
	Jožef Stefan Institute, Jamova c. 39, 1000 Ljubljana, Slovenia}
	\affiliation{Faculty of Mathematics and Physics, University of Ljubljana, Jadranska u. 19, 1000 Ljubljana, Slovenia}
\author{Kwang-Yong Choi}
\email{choisky99@skku.edu}
\affiliation{Department of Physics, Sungkyunkwan University, Suwon 16419, Republic of Korea}
\author{P. Khuntia}
\email[]{pkhuntia@iitm.ac.in}
\affiliation{Department of Physics, Indian Institute of Technology Madras, Chennai 600036, India}
\affiliation{Quantum Centre of Excellence for Diamond and Emergent Materials, Indian Institute of Technology Madras,
	Chennai 600036, India.}

\date{\today}

\begin{abstract}

	 Quantum magnets based on honeycomb lattices with low-coordination number offer a viable ground to realize  exotic emergent quantum  excitations and phenomena arising from the interplay between competing magnetic interactions, spin correlations, and spatial anisotropy. However, unlike their low-spin analogues, high-spin honeycomb lattice antiferromagnets have remained comparatively less explored in the context of capturing the classical analogs of quantum phenomena.
	Herein, we report the crystal structure, magnetic susceptibility,  specific heat, and electron spin resonance (ESR), complemented by \textit{ab initio} density functional theory (DFT) calculations on polycrystalline samples of FeP$_{3}$SiO$_{11}$ wherein the Fe$^{3+}$ ions decorate a nearly-perfect $S$ = 5/2 honeycomb lattice without any site disorder among constituent atoms. Above 150 K, an antiferromagnetic Weiss temperature $\theta_{\rm CW}$ = $-$ 12 K is observed consistent with DFT calculations, which suggest the presence of strong intra-planar nearest-neighbor and weaker inter-planar further nearest-neighbor exchange interactions.  An anomaly at $T_{N}$ = 3.5 K in specific heat and magnetic susceptibility reveals the presence of a long-range  ordered ground state in zero field. Above $T_{N}$, ESR evidences short-range spin correlations and unsaturated magnetic entropy, while below  $T_{N}$ unconventional excitations are seen via power-law specific heat. A spin-flop transition is observed at an applied field of $\mu_0 H_{c1} = 0.2$~T. At higher applied fields, $T_{N}$ is gradually suppressed down to zero at $\mu_{0}H_{\rm c2}$ = 5.6 T with a 2D critical exponent $\beta = 0.255$. Above  $\mu_{0}H_{\rm c2}$,  a broad maximum in specific heat due to gapped magnon excitations indicates the emergence of an interesting nearly-polarized state dressed by a disordered state in the honeycomb lattice antiferromagnet FeP$_{3}$SiO$_{11}$.
\end{abstract}
\maketitle
\section{Introduction}
Quantum materials, wherein the interplay of electron correlations, competing interactions between local moments, entanglement, and topology of the underlying electronic band structure converge, offer a promising framework for realizing novel quantum states that go beyond the Landau paradigm of symmetry breaking \cite{Balents2010}. One such elusive state is a quantum spin liquid (QSL), wherein strong quantum fluctuations induced by frustrated or competing interactions prohibit the development of classical ordered Ne\'el states with conventional spin-wave excitations \cite{Savary_2016,KHATUA20231}. QSLs are characterized by  non-local order parameters and fractionalized excitations, such as spinons or Majorana fermions,  which are believed to emerge from the quantum entangled state of local moments in insulators and hold promise for topological quantum computation \cite{ANDERSON1973153,Takagi2019,RevModPhys.80.1083}.\\
Beyond the geometrically-frustrated two-dimensional triangular and kagome lattices \cite{KHATUA20231,Khuntia2020,jeon2023oneninth}, contemporary investigations on quantum materials are increasingly focused towards experimental realizations of QSLs in bipartite honeycomb lattices \cite{TREBST20221}. Theoretical studies suggest that the nearest-neighbor isotropic exchange interactions on a honeycomb lattice, which are free from
frustration, lead to a long-range ordered state \cite{Fouet2001,PhysRevB.107.064409}. However, additional competing further-neighbor interactions, which introduce frustration-induced quantum fluctuations, can lead to exotic quantum states, including the QSL state \cite{Meng2010,PhysRevLett.113.027201,PhysRevB.83.144414,doi:10.1143/JPSJ.79.114705,katsura1986ground,PhysRevB.105.174403}.
Recently, a few promising honeycomb
lattice antiferromagnets have been identified, including InCu$_{2/3}$V$_{1/3}$O$_{3}$ ($S$
= 1/2) \cite{PhysRevB.100.144442,KATAEV2005310}, $A_{3}$Ni$_{2}$SbO$_{6}$ ($A$ = Li, Na with $S$ = 1) \cite{PhysRevB.96.024417,WERNER2019100}, Ag$_{3}$LiMn$_{2}$O$_{6}$ ($S$ = 3/2) \cite{PhysRevB.99.144429}, and
Bi$_{3}$Mn$_{4}$O$_{12}$(NO$_{3}$)($S$ = 3/2) \cite{PhysRevB.85.184412,PhysRevLett.105.187201}, where isotropic Heisenberg interaction between 3$d$ magnetic moments stabilizes a long-range ordered state at low temperatures. In addition, these materials have been studied to elucidate the impact of quantum effects, stemming from additional further-neighbor Heisenberg interactions, to understand the physics of QSL,  field-induced magnetic order and topological phases such as the Berezinskii-Kosterlitz-Thouless phase \cite{PhysRevB.86.140401,PhysRevB.96.140404,geng2020distant,PhysRevB.99.144429}.\\
To further understand the  quantum effects on the simplest isotropic two-dimensional honeycomb lattices, recent studies have focused on rare-earth-based magnets  as an alternative to transition metal-based systems with pure $S$ = 1/2 moments \cite{Arh2022}. The rare-earth based honeycomb lattices  \cite{Sala2023,PhysRevB.102.014427,Wessler2020,PhysRevB.108.054442}  with $J_{\rm eff}$ = 1/2 moments of magnetic Kramers ions have drawn significant attention to explore quantum fluctuations governed by low dimensionality, and low
connectivity in the isotropic Heisenberg model on honeycomb lattice. Remarkably, YbCl$_{3}$ wherein the rare-earth Yb$^{3+}$ ions form a honeycomb lattice exhibits characteristic features  of a two-dimensional (2D)  Heisenberg antiferromagnet with a potential to realize exotic   quantum phenomena including quantum Bose gas near the critical field \cite{matsumoto2023quantum}, Van Hove singularities \cite{Sala2021}, and non-trivial field-induced magnetic excitations \cite{Sala2023}. Interestingly, the observation of continuous excitations in another rare-earth-based honeycomb lattice antiferromagnet YbBr$_{3}$ points towards collective quantum behavior of local moments in the plaquette phase which was believed to exist between the N\'eel order and columnar order in the isotropic Heisenberg model \cite{Wessler2020,PhysRevLett.113.027201}.\\ 
The frustration-induced quantum fluctuations owing to further-neighbor exchange interaction offers a promising route for the experimental realization of a QSL state in the isotropic Heisenberg model on a honeycomb lattice \cite{PhysRevB.96.094432}. Remarkably, an exactly solvable Kitaev model for $S$ = 1/2 spins on a 2D honeycomb lattice was proposed by Alexei Kitaev to demonstrate a  QSL state \cite{KITAEV20062}.  In this scenario, the spin frustration induced bond-dependent anisotropic Kitaev interactions between local moments on the honeycomb lattice accounts for the QSL state with fractionalized Majorana fermion and gauge flux excitations. In this context, the 5$d$/4$d$ transition metal-based edge-sharing  honeycomb octahedral lattices of magnetic ions, that favor
nearly 90$^{\circ}$ metal–ligand–metal bond angles for electron hopping paths,  have been studied widely to understand the ramifications of Kitaev interactions in  macroscopic observable as well as in low-energy excitation spectra \cite{PhysRevLett.108.127204,Do2017,Banerjee2018,Park_2024,Lin2021,PhysRevB.103.214447,PhysRevLett.124.087205,Jin2022}.
Despite enormous efforts, the ground state of the existing Kitaev
magnet remains intriguing, mainly due to the presence of finite Heisenberg exchange, additional nearest-neighbor interactions, and structural imperfections \cite{TREBST20221}.\\
In comparison to the honeycomb lattices based on $S = 1/2$, the realization of spin-5/2 honeycomb lattices is relatively scarce.  Interestingly,  frustrated Heisenberg  nearest-and next-nearest-neighbor interaction have been proposed to lead to an exotic spiral spin liquid state in $S$ = 5/2  honeycomb lattice antiferromagnet FeCl$_{3}$, which, however, exhibits a magnetic phase transition at low temperatures \cite{PhysRevLett.128.227201}. The realization of a spiral spin liquid with emergent fractional excitations on a honeycomb lattice is promising research direction.  Moreover, high-spin honeycomb lattices offer the opportunity to explore classical spin liquid states \cite{PhysRevB.96.134408} above the transition temperature, similar to observations in frustrated kagome lattice  Li$_{9}$Fe$_{3}$(P$_{2}$O$_{7}$)$_{3}$(PO$_{4}$)$_{2}$ ($S=$ 5/2) \cite{PhysRevLett.127.157202,Rousochatzakis2018}.  In addition to the edge-sharing octahedra with $90^{\circ}$ metal-ligand-metal bond angle, the possibility of Kitaev interaction through extended superexchange pathway between discrete octahedra of magnetic ions has been proposed in a honeycomb lattice RuP$_{3}$SiO$_{11}$ \cite{abdeldaim2024kitaev}.  \\  
In certain magnetic materials, the presence of competing magnetic interactions can lead to a phenomenon where only a small fraction of spins stabilize long-range order, while the remaining spins  maintain in a fluctuating
state. The ground state of such frustrated magnets can be tuned by external perturbations such as a magnetic field, to realize a field-induced spin liquid ground state just before reaching the fully polarized state. Experimental evidence of a field-induced spin-liquid like state has been observed in several promising honeycomb lattice-based  QSL candidates such as Na$_{2}$Co$_{2}$TeO$_6{}$ \cite{Lin2021}, $\alpha$-RuCl$_{3}$ \cite{PhysRevLett.119.037201,Li2021}, and YbCl$_{3}$ \cite{matsumoto2023quantum}. \\
Detailed studies distinguishing the signature of continuum excitations in QSL states where electron spins are entangled from those in ordered magnets are lacking. Furthermore,  
to investigate whether field-induced phenomena characteristic of low-spin systems persist in high-spin honeycomb lattices, we focus on an unexplored honeycomb-lattice material FeP$_{3}$SiO$_{11}$ (henceforth, FPSO) \cite{PhysRevB.98.045121}. Here, Fe$^{3+}$ ($S$ = 5/2) ions form  a perfect 2D honeycomb lattice perpendicular to the $c$-axis without any statistical distribution between the crystallographic sites. The magnetic susceptibility data show an anomaly at $\sim$ 3.5 K indicating the presence of a phase transition in this honeycomb lattice. The obtained negative Weiss temperature ($\theta_{\rm CW}$ $\approx$ $-$12 K) from the fit of  high-temperature magnetic susceptibility with Curie-Weiss law indicates the presence of dominant antiferromagnetic interactions between $S$ = 5/2 spins. A $\lambda$-type anomaly in specific heat reveals  the emergence of a long-range magnetic ordered state below the N\'eel temperature $T_{N}$ = 3.5 K, consistent with magnetic susceptibility data. At $T \ll T_{\rm N}$, the power-law behavior of magnetic specific
heat in a zero field is ascribed to  unconventional magnetic excitations. Tuning of the magnetic ground state is evidenced by a shift of $T_{N}$ towards lower temperatures in an applied magnetic field. The complete suppression of the 3.5 K anomaly and the appearance of a double-peak-like maximum in the specific heat in a magnetic field of 6 T indicate the presence of a disordered state on top of a nearly-polarized state above a critical field $\mu_{0}H_{c2} = 5.8$ T. The critical behavior of the ESR linewidth, along with a few percent of unsaturated magnetic entropy, provides the evidence of short-range spin correlations above $T_{N}$ in the present honeycomb lattice  antiferromagnet. 
In addition to the intra-planar nearest-neighbor exchange interactions, the further-neighbor inter-planar exchange interaction, confirmed by DFT calculations, establishes a long-range ordered state in FPSO.
\begin{figure*}
 	\centering
 	\includegraphics[width=\textwidth]{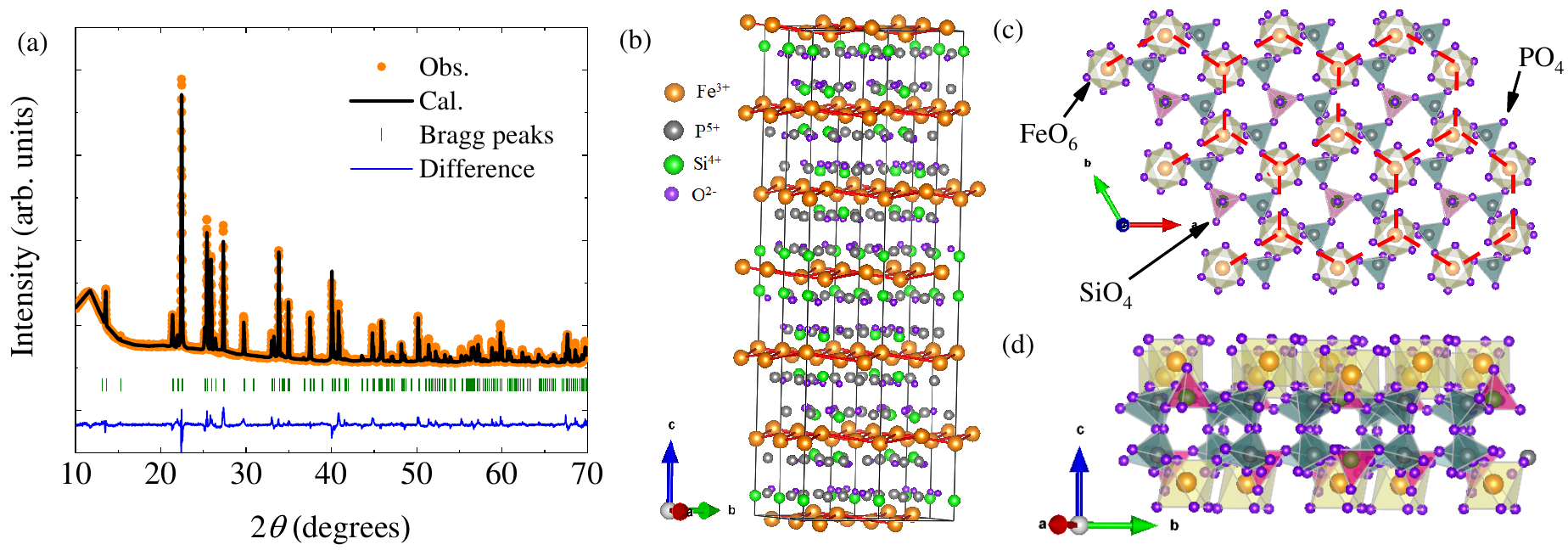}
 	\caption{(a) Rietveld refinement profile of powder x-ray diffraction data. The filled orange circles and the solid black line represent the experimental data and the calculated diffraction pattern, respectively. Olive vertical bars represent allowed Bragg positions, and the solid blue line at the bottom is the difference between the experimental and calculated diffraction intensity. (b) Schematic of one unit cell of FeP$_{3}$SiO$_{11}$, wherein the honeycomb layers are stacked perpendicular to the $c$-axis. (c) Top view of the intra-planar FeO$_{6}$ octahedra which are  connected with nearest-neighbor FeO$_{6}$ octahedra via PO$_{4}$ tetrahedra and form a superexchange path  Fe--O--P--O--Fe in the $ab$-plane. Magnetic Fe$^{3+}$ ions with $S$ = 5/2 constitute a honeycomb lattice, as indicated by the dotted lines.  (d) Parallel to the $c$-axis, the FeO$_{6}$ octahedra are connected through P$_{2}$O$_{7}$ and form a inter-planar interaction path  Fe--O--P--O--P--O--Fe.  }{\label{FPSO1}}.
 \end{figure*}
\begin{table}
	\caption{\label{table} Structural parameters of FeP$_{3}$SiO$_{11}$ from a  Rietveld refinement of x-ray diffraction data at 300 K. (Space group: $R \bar{3}c$, $a = b$ = 8.29 \AA , $c$ = 39.054 \AA, $\alpha = \beta = 90^{\circ}$, $ \gamma$ = 120 $^\circ$
		and $\chi^{2}$ = 2.86, R$_{\rm wp}$ = 5.29 \text{\%}, R$_{\rm p}$ = 3.56 \text{\%}, and R$ _{\rm exp}$ = 1.84\text{\%}.)}
	\begin{tabular}{c c c c c  c c} 
		\hline \hline
		Atom & Wyckoff position & \textit{x} & \textit{y} &\textit{ z}& Occ.\\
		\hline 
		Fe & 12$c$ & 0 \ \ & 0 \ \ & 0.157 \ \ & 1 \\
		P& 36$f$ &  0.369 & 0.037 & 0.121 &   1 \\
		Si &12$c$ & 0 & 0 & 0.04 & 1\\
		O$_{1}$& 36$f$ & 0.048 & 0.197 & 0.051 & 1 \\
		O$_{2}$ & 36$f$ & 0.28 & 0.075 & 0.127 & 1\\
		O$_{3}$ & 36$f$ & 0.142 & 0.228 & 0.191 & 1\\
		O$_{4}$ & 18$e$&  0.209 & 0 & 0.75 & 1 \\
		O$_{5}$ & 6$b$& 0 & 0 & 0 & 1\\	
		\hline
	\end{tabular}
\end{table} 
\section{Experimental details} Polycrystalline samples of FPSO employed in this study were synthesized by a two-stage preparation procedure. Initially, the precursor Fe(PO$_{3}$)$_{3}$ was synthesized using a conventional solid-state reaction method, with H$_{3}$PO$_{4}$ (98\text{\%}, Alfa Aesar) and Fe$_{2}$O$_{3}$ (99.998\text{\%},  Alfa Aesar) as the starting materials. The stoichiometric mixtures were formed into pellets and then sintered at 300$^\circ$C for 20 hr in an alumina crucible in air, with intermediate steps, until a single phase was achieved at 550$^\circ$C for 30 hr \cite{Zhou2009}. In a subsequent stage, the FPSO compound was synthesized from a stoichiometric mixture of the Fe(PO$_{3}$)$_{3}$ precursor and SiO$_{2}$ (99.999\text{\%}; Alfa Aesar). The mixtures were pelletized and introduced into an evacuated and sealed quartz tube, followed by sintering at 500$^\circ$C for 24 hr in air. 
Several intermediate temperature stages were iterated before reaching a single phase at 800$^\circ$C for 48 hr \cite{ELBOUAANANI1999565}.\\
Powder x-ray diffraction (XRD)
data were collected using a smartLAB Rigaku x-ray diffractometer with Cu K$_{\alpha}$ radiation ($\lambda $ = 1.54 {\AA}) at room temperature. Magnetization measurements were performed using
the VSM option of a Physical Properties Measurement System (PPMS, Quantum Design)  in a temperature range of 2 K $\leq$ \textit{T} $\leq$ 300 K and in magnetic fields up to 9 T. Specific heat measurements were performed  using a PPMS by a thermal relaxation method in a temperature range of 1.9 K $\leq$ \textit{T} $\leq$ 250 K  in magnetic fields up to 9 T.\\
ESR measurements were performed at X-band ($\nu$ =  9.5 GHz)  using  a Bruker EMXplus-9.5/12/P/L  spectrometer with  a continuous He flow cryostat in a temperature range of 4.5 K $\leq$ $T$ $\leq$ 300 K. 
\begin{figure*}[t]
	\centering
	\includegraphics[width=\textwidth]{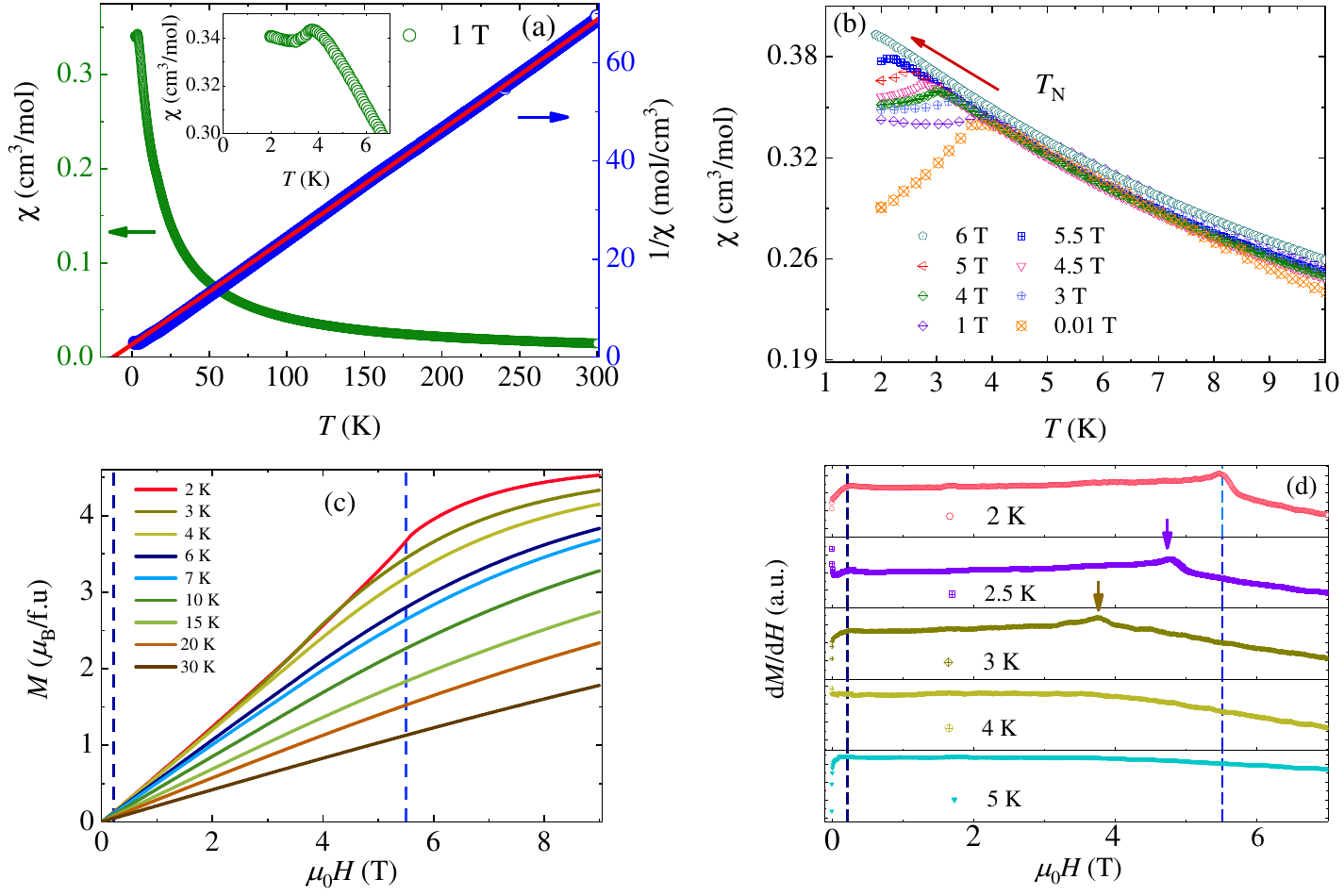}
	\caption{(a) Temperature dependence of the magnetic susceptibility, $\chi(T)$, and the inverse magnetic susceptibility of FeP${_3}$SiO$_{11}$ are shown on the left and right $y$-axes, respectively, in a magnetic field of $\mu_{0}H$ = 1 T. The inset shows a magnified view of an anomaly at $T_{N}$ $\approx$ 3.8 K in the temperature dependence of $\chi(T)$ data. The red line is a  Curie-Weiss fit to the  high-temperature inverse susceptibility data.   (b) Temperature dependence of $\chi(T)$ in the temperature range 2 K $\leq$ $T$ $\leq$ 10 K in several magnetic fields upto 6 T. The brown arrows shows the direction of the change of $T_{N}$ as the magnetic field increases.  (c) Magnetization, $M$, and (d) differential magnetic susceptibility, $\mathrm{d}M /\mathrm{d}H$, as a function of  external magnetic field at several temperatures. Vertical dashed lines denote  changes in the slope of the 2 K magnetization curve, occurring at $\mu_{0}H_{\rm c1}$ = 0.2 T and $\mu_{0}H_{\rm c2}$ = 5.6 T, respectively. The shift of peak-like feature towards higher fields as the temperature approaches $T_{N}$ is indicated by an arrow.   }{\label{FPSO2}}.
\end{figure*}
\section{Results and discussion}
\subsection{Rietveld refinement  and crystal structure } 
Phase purity of  the polycrystalline samples of FPSO was confirmed from a Rietveld refinement of powder XRD data  using GSAS software \cite{Toby:hw0089}. 
The proposed structural model in Ref.~\cite{ELBOUAANANI1999565} based on neutron diffraction data on FPSO was used as a starting guess. The refined structural parameters are summarized in Table~\ref{table}.  The Rietveld refinement profile of XRD data 
in Fig.~\ref{FPSO1}(a) indicates that FPSO crystallizes in a trigonal space group ($R\bar{3}c$) with no detectable statistical distribution between atomic sites. All the peaks observed in the XRD pattern are accurately indexed to the expected nuclear Bragg peaks, indicating the lack of any detectable secondary phase in FPSO.\\  One unit cell of FPSO is shown in Fig.~\ref{FPSO1}(b), wherein the nearest-neighboring Fe$^{3+}$ ions ($\approx$ 4.842(1) Å) of  form two-dimensional ($ab$-plane) honeycomb lattices stacked along the $c$-axis. The structure of FPSO consists of a three-dimensional network of corner-sharing FeO$_{6}$, P$_{2}$O$_{7}$, and Si$_{2}$O$_{7}$ units in a trigonal unit cell. The intra-planar FeO$_{6}$ octahedra are linked together through the PO$_{4}$ tetrahedra, forming a Fe--O--P--O--Fe bridge for the first nearest-neighbor intra-planar exchange interaction [see Fig.~\ref{FPSO1}(c)].
On the other hand, inter-planar FeO$_{6}$ octahedra are connected via P$_{2}$O$_{7}$ double tetrahedra along the $c$-axis, establishing a Fe--O--P--O--P--O--Fe exchange path for inter-planar exchange interactions  [see Fig.~\ref{FPSO1}(c)]. The Si$_{2}$O$_{7}$ double tetrahedra located at the center of the hexagon share corners with the nearest PO$_{4}$ tetrahedra, resulting in an intra-planar second nearest-neighbor exchange path  Fe--O--P--O--Si--O--P--O--Fe [see Fig.~\ref{FPSO1}(d)]. \\ 
 From the structural parameters, it is observed that the honeycomb layers are separated by an inter-planar distance of 7.248(3) Å, which is shorter than the intra-planar second nearest-neighbor distance of 8.289(2) Å. One would thus expect that the ground state of the present honeycomb lattice can be understood by the combination of intra-planar first nearest-neighbor interaction and inter-planar interactions. The intra-planar second nearest-neighbor interaction, which typically introduces spin frustration, is expected to be very weak. Interestingly, our DFT calculations strongly support (see subsection \ref{subsec:dft}) this scenario, revealing that the intra-planar nearest-neighbor exchange interaction $J$ is approximately three times greater than the inter-planar magnetic exchange interaction, while intra-planar second nearest-neighbor interactions are negligible. \begin{figure}
 	\centering
 	\includegraphics[width=0.9\columnwidth]{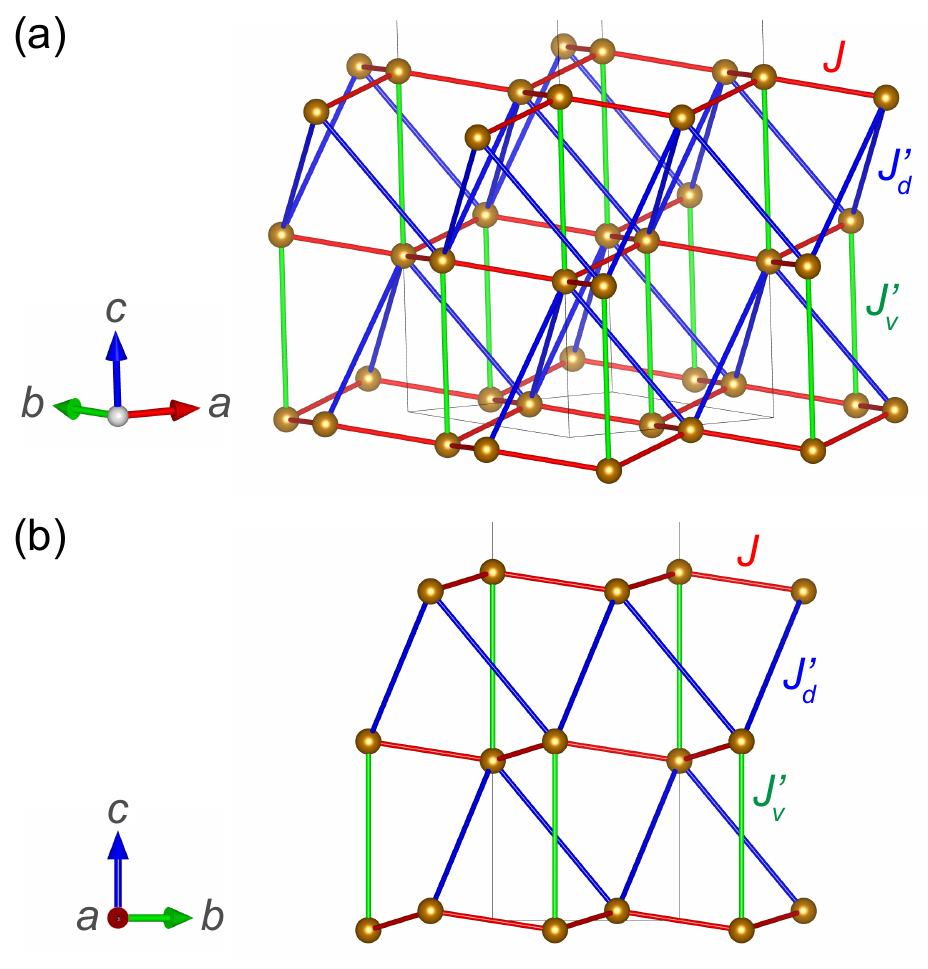} 
 	\caption{%
 		Isotropic spin model of FeP$_{3}$SiO$_{11}$ from \textit{ab initio} DFT calculations. 
 		Each Fe$^{3+}$ spin is connected to three in-plane nearest-honeycomb-neighbouring Fe$^{3+}$ spins via $J$ (red), one Fe$^{3+}$ spin directly above/below it along the $c$ axis in an adjacent honeycomb layer via $J'_v$ (green), and three non-neighbouring Fe$^{3+}$ spins from a honeycomb hexagon in the other adjacent honeycomb layer below/above it along the $c$ axis via $J'_d$ (blue). 
 		Shown is a $2 \times 2 \times 1/3$ super-/sub-cell of the unit cell of FeP$_{3}$SiO$_{11}$ viewed along (a) an arbitrary direction, and (b) along the $a$ axis.}
 	\label{fig_spin_model}
 \end{figure}
\subsection{Magnetic susceptibility}
Figure \ref{FPSO2}(a) depicts the temperature dependence of magnetic susceptibility $\chi(T)$ in a magnetic field of $\mu_{0}H$ = 1 T. As the temperature decreases,  $\chi(T)$ exhibits a gradual increase and reveals an anomaly around  $T_{N}$ $\approx$ 3.8 K, which indicates the onset of long-range magnetic order in FPSO [see also Fig.~\ref{FPSO2}(b)]. To estimate the dominant magnetic interactions between $S$ = 5/2 spins of Fe$^{3+}$ ions,  inverse magnetic susceptibility data 1/$\chi(T)$ were fitted  to the Curie--Weiss (CW) law,  $\chi = \chi_{0}$ + $C$/($T$ $-$ $\theta_{\rm CW}$), at $T > 150$~K. Here, $\chi_{0}$ = $-$ 4.068 $\times 10^{-4}$ $\mathrm{cm}^3/\mathrm{mol}$  is the sum of temperature-independent core diamagnetic susceptibility ($\chi_{\rm core}$) and Van Vleck paramagnetic susceptibility ($\chi_{\rm VV}$), $C$ = $ 4.66$ $\mathrm{cm}^3 \mathrm{K}/\mathrm{mol}$ is the Curie constant and  $\theta_{\rm CW}$ = $-12.0(1) \mathrm{K}$ is the Weiss temperature which provides a measure of average magnetic exchange interaction between  magnetic ions.  The  effective magnetic moment was determined to be $\mu_{\rm eff}$ = $\sqrt{8C}$= 6.10 $\mu_{\rm B}$,  which is close to
the value $g \sqrt{S(S+1)}$ = 5.92 $\mu_{\rm B}$ , where $g = 2$, expected for $S$ = 5/2 moments of free Fe$^{3+}$ ions according to Hund’s rule, consistent with paramagnetic behavior at these high temperatures.
The negative $\theta_{\rm CW}$ indicates the presence of dominant antiferromagnetic interactions between the $S$ = 5/2 spin of Fe$^{3+}$
ions. \\ 
To better understand the ground state  of FPSO under external tuning parameters, $\chi(T)$ measurements were performed  as a function of temperature in several magnetic fields as shown in Fig.~\ref{FPSO2}(b). Interestingly, we observed
that upon increasing magnetic field, the $T_{N}$ shifts toward lower temperatures. Surprisingly,
the long-range antiferromagnetic order is completely suppressed in magnetic fields  $\mu_{0}H\geq$ 6 T. The shift of $T_{N}$ toward lower temperatures on
increasing the field strength is typical for antiferromagnetically ordered state \cite{abdeldaim2024kitaev}.
\\
Figure~\ref{FPSO2}(c) shows the magnetization $M(H)$ as a function of applied magnetic field $\mu_0 H$ at several temperatures. At the lowest measured temperature of 2 K, a noticeable change in the slope of the magnetization curve is seen at two different magnetic field values, specifically at $\mu_{0}H_{\rm c1}$ = 0.2 T and $\mu_{0}H_{\rm c2}$ = 5.6 T. This is indicated by vertical dashed lines in Fig.~\ref{FPSO2} (c,d) for 2 K. To better characterize this change in slope, the corresponding differential magnetic susceptibility $dM/dH$ was calculated, as depicted in Fig.~\ref{FPSO2}(d). Clear peak-like features are observed at $\mu_{0}H_{\rm c1}$ = 0.2 T and $\mu_{0}H_{\rm c2}$ = 5.6 T in  $dM/dH$. The peak-like anomalies in $dM/dH$, which appear below $T_{N}$, gradually diminish with increasing temperature and eventually vanish above $T_{N}$.
  It is worth to noting that at $\mu_{0}H = $ 6 T, which is close to $\mu_{0}H_{\rm c2}$, the anomaly in $\chi(T)$ disappears. On the other hand, the anomaly at $\mu_{0}H_{\rm c1}$ = 0.2 T can be attributed to a spin-flop-like transition \cite{doi:10.7566/JPSJ.86.074706}. A similar scenario of a field-induced phase is also observed in a few promising   honeycomb lattice antiferromagnets \cite{doi:10.7566/JPSJ.86.074706,PhysRevMaterials.3.124406}.   Furthermore, the finite slope of $M(H)$ above $\mu_{0}H$ = 6 T implies the presence of appreciable magnetic anisotropies \cite{RODRIGUEZ201639,coey2010magnetism}.
  \subsection{\textit{Ab initio} DFT magnetic exchange calculations}\label{subsec:dft}
  Exchange couplings in FPSO were calculated using the \mbox{CASTEP} plane-wave \textit{ab initio} density functional theory (DFT) code~\cite{clark2005first} using ultrasoft pseudopotentials and the LDA exchange--correlation functional with additional Hubbard repulsion in an LSDA+$U$ scheme~\cite{anisimov1997first}. 
  The LDA functional was chosen because it was reported to generally better reproduce experimental data on iron-bearing minerals~\cite{hsu2011hubbard} than the alternative PBE exchange--correlation functional~\cite{perdew1996generalized}, and since it is expected to be generally more appropriate for exchange-coupling calculations than the PBE functional~\cite{sharma2018source}.
  An effective additional on-site Hubbard repulsion of $U_\mathrm{eff} = U - J_H = $ 4 eV, where $U$ is the bare Hubbard repulsion and $J_H$ is Hund's coupling, was chosen for Fe$^{3+}$ $3d$ orbitals throughout, consistently with DFT results on the $S = 5/2$ zigzag-chain material Ba$_{5}$Fe$_{2}$ZnIn$_{4}$S$_{15}$~\cite{almoussawi2022preparation} and Fe$_{3}$O$_{4}$(001)~\cite{novotny2013probing}. 
  Calculations were carried out on a DFT geometry-optimized unit cell of FPSO with a 1200 eV plane-wave energy cutoff and a $3 \times 3 \times 1$ Monkhorst-Pack grid reciprocal-space sampling~\cite{monkhorst1976special} to achieve numerical convergence. 
  All calculations were converged to within a total energy tolerance of 0.1 neV per atom in the self-consistent field (SCF) DFT loop, while geometry-optimization calculations were converged to within a 0.05 eV $\text{\AA}^{-1}$ force tolerance on atoms.\\  Isotropic (Heisenberg) exchange couplings were extracted within the total-energy (broken-symmetry) DFT framework~\cite{riedl2019ab} by calculating total DFT energies of $119$ random collinear spin configurations with maximal spins $S = 5/2$ pointing along the crystallographic ${\pm}c$ axis. Extracted exchange strengths were corrected by taking into account the fact that the Noodleman broken-symmetry-singlet state $\ket{+S, -S}$ and the maximal-spin state $\ket{+S,+S}$ have an expected energy difference of $J/[1+1/(2S)]$ for general $S$ and an isotropic spin Hamiltonian, $\mathcal{H}$ = $J \mathbf{S}_1 \cdot \mathbf{S}_2$.
  For $S = 1/2$ this would reproduce the standard broken-symmetry energy difference $J/2$~\cite{riedl2019ab}, while for $S = 5/2$ we obtain a renormalized broken-symmetry effective energy difference of $5J/6$. \\ 
  Our calculations yield the dominant in-plane nearest-neighbor antiferromagnetic honeycomb exchange interaction $J = 0.860(5)$ K (Fe--Fe distance ${4.8}$  $\text{\AA}$), a vertical antiferromagnetic interlayer exchange interaction $J'_v = 0.165(5)$ K (Fe--Fe distance 7.2 $\text{\AA}$), and a diagonal antiferromagnetic inter-layer exchange interaction $J'_d =$ 0.240(5) K (Fe--Fe distance 7.5 $\text{\AA}$), 
  yielding a total \textit{ab initio} antiferromagnetic Weiss temperature of $\theta_\mathrm{CW} = -10.1$ K, which is broadly consistent with experimental results. 
  All other allowed exchange interaction were found to be smaller than $20$ mK. Normalizing the calculated exchanges by the ratio between the experimental and calculated Weiss temperature, we obtain the final estimates of exchanges $J$, $J_v'$, and $J_d'$ of $1.02(1)$, $0.196(6)$, and $0.285(6)$~K, respectively. 
  The resulting 3D spin model is presented in Fig.~\ref{fig_spin_model}.  
\begin{figure*}
	\centering
	\includegraphics[width=\textwidth]{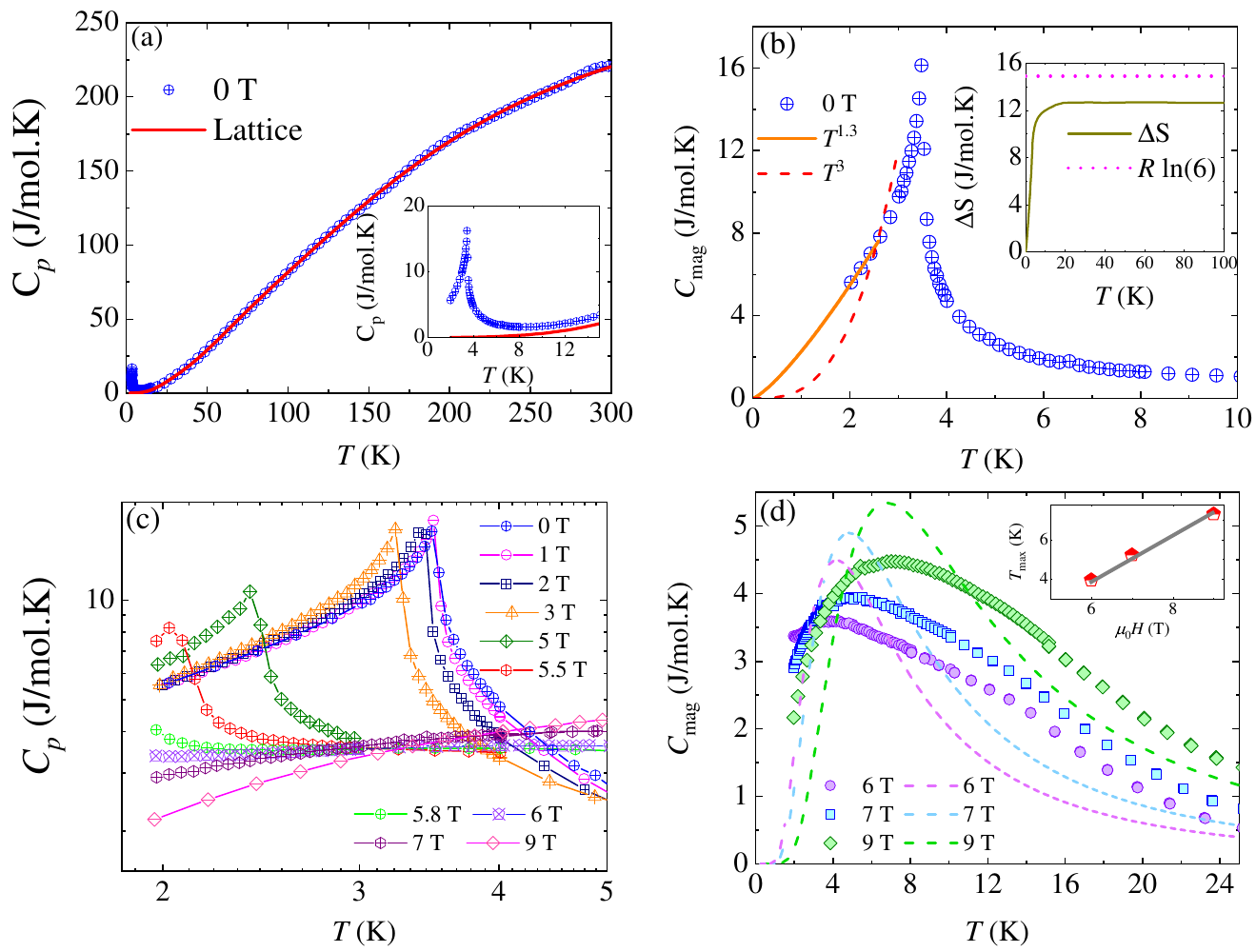}
	\caption{(a) Temperature dependence of specific heat \textit{C}$_{\rm p}$ of  FeP$_{3}$SiO$_{11}$  in zero-magnetic field. The solid red line depicts the fitted lattice contribution to $C_{p}$ via Eq.\ref{eqn:debye}. The inset shows a zoom-in look at the low-$T$ $C_p$ anomaly. (b) 
		Temperature dependence of magnetic specific heat ($C_{\rm mag}$) in zero-magnetic field in the temperature range 2 K $\leq$ $T$ $\leq$ 10 K. The orange solid line indicates a power-law fit as described in the main text while the dashed red line corresponds to $T^{3}$
		behavior. Inset shows the estimated entropy release $\Delta S$ = $\int (C_{\rm mag}/T) dT$ as a function of temperature where the horizontal pink line is the expected entropy of $R$ln(6) for $S = 5/2$ spin of Fe$^{3+}$ ions.  (c) Temperature dependence of $C_{p}$ in the temperature range 2 K $\leq$ $T$ $\leq$ 5 K in several magnetic fields on a log--log scale.  (d) Temperature dependence of $C_{\rm mag}$ in high magnetic fields. Dashed curves represent
		 fits with a two-level Schottky model as described in the text. The inset depicts the variation of  the broad maximum with the applied magnetic field.
	 }{\label{FPSO3}}.
\end{figure*} 
\subsection{Specific heat}
Specific heat is an excellent probe for unveiling  the ground state
and the nature of low-lying excitations in frustrated magnets. In order to confirm the presence of long-range magnetic  order in FPSO, specific heat measurements were performed in several magnetic fields down to 2 K. The temperature dependence of specific heat $C_{p}$ in zero field is shown in Fig.~\ref{FPSO3}(a). It exhibits a $\lambda$-like anomaly at  $T_{N}$ = 3.5 K, which  confirms the presence of long-range magnetic order in FSPO, consistent with $\chi( {T})$ data. A slight discrepancy between the temperature where the anomaly appears in specific heat and magnetic susceptibility data is observed, similar to the other honeycomb lattice materials \cite{PhysRevB.99.144429}. One possible reason for this inconsistency could be subtle variations in the experimental conditions, leading to differences in the sensitivity of the measurements to the underlying magnetic behavior. Henceforth, we consider the anomaly at $T_{N}$ = 3.5 K in the zero-field specific heat data as the actual transition temperature in FPSO, as specific heat is a highly sensitive technique for tracking phase transitions in correlated quantum materials.   \\ 
In insulators based on transition metals like FPSO, the measured $C_{p}$ comprises the combined effects of lattice-specific heat due to phonons ($C_\mathrm{latt}$) and magnetic-specific heat originating from the $S = 5/2$ spins of Fe$^{3+}$ magnetic ions ($C_\mathrm{mag}$). In the absence of a suitable-non magnetic analog of FPSO, the lattice contribution to the $C_{p}$ data was fit with a combination of one Debye and three Einstein functions  in the temperature range 70 K $\leq$ $T$ $\leq$ $250$ K  [solid line on Fig.~\ref{FPSO3}(a)]  \cite{kittel2021introduction}
\begin{equation*}
C_{\rm latt}(T)=C_{D}[9k_{B} \left(\frac{T}{\theta_{D}}\right)^{3}\int_{0}^{\theta_{D}/T}\frac{x^{4}e^{x}}{(e^{x}-1)^{2}}dx]
\end{equation*}
\begin{equation}\label{eqn:debye}
+\sum_{i=1}^{3} C_{E_{i}}[3R\left(\frac{\theta_{E_i}}{T}\right)^{2}\frac{\text{exp}(\frac{\theta_{E_{i}}}{T})}{(\text{exp}(\frac{\theta_{E_{i}}}{T})-1)^{2}}], 
\end{equation}
 where $\theta_{D}$ = 148(12) K is the Debye temperature, $\theta_{E_1} = 140(8)$, $\theta_{E_2} = 304(15)$, $\theta_{E_3} = 723(18)$~K are the Einstein temperatures of the three optical phonon modes, \textit{R} and \textit{k}$_{B}$ are the molar gas and Boltzmann constant, respectively.   To minimize the fitting parameters, a constant ratio between the coefficients $C_{D}$ and $ C_{E_i}$ was assigned, accounting for the ratio of heavy atoms (Fe, P, Si) to light atoms (O) in FPSO. Upon lowering the temperature, the measured $C_{p}$ begins to diverge from the lattice fit below 15 K, as illustrated in the inset of Fig.~\ref{FPSO3}(a). This implies that below 15 K, significant magnetic correlations start to develop, consistent with magnetic susceptibility results. \\
To estimate the thermal evolution of entropy release associated with spin dynamics, the temperature dependence of $C_{\rm mag}$ was obtained after subtracting $C_{\rm latt}$ and is shown in Fig.~\ref{FPSO3}(b) in zero field.  It exhibits a clear $\lambda$-like anomaly at $T_{N}$, a hallmark of a thermodynamic phase transition,  suggesting the appearance of long-range magnetic order owing to finite inter-planar exchange interactions in addition to intra-planar exchange interaction in FPSO. At  $T$ $\ll$ $T_{N}$, long-rage-ordered antiferromagnets with conventional magnon excitations are expected to exhibit $C_\mathrm{mag} \propto T^3$ behavior \cite{Bloch1930}. However, in FPSO we observe $C_{\rm mag}  \propto T^{1.3}$ power-law behavior in zero field [Fig.~\ref{FPSO3}(b)], suggesting the presence of unconventional magnetic excitations in its ground state \cite{PhysRevLett.131.146701}. Namely, the presence of such behavior can be connected to a  unconventional spin excitations below $T_{N}$ due to the presence of significant competing interactions  \cite{PhysRevLett.127.157204,doi:10.1126/science.aah6015}. \\In FPSO, the value of the empirical frustration parameter $f$ = $|\theta_{\rm CW}|$/$T_{N}$ $\approx$ 4 indicates the existence of moderate spin frustration resulting from competing magnetic interactions.
\begin{figure}
	\includegraphics[width=0.45\textwidth]{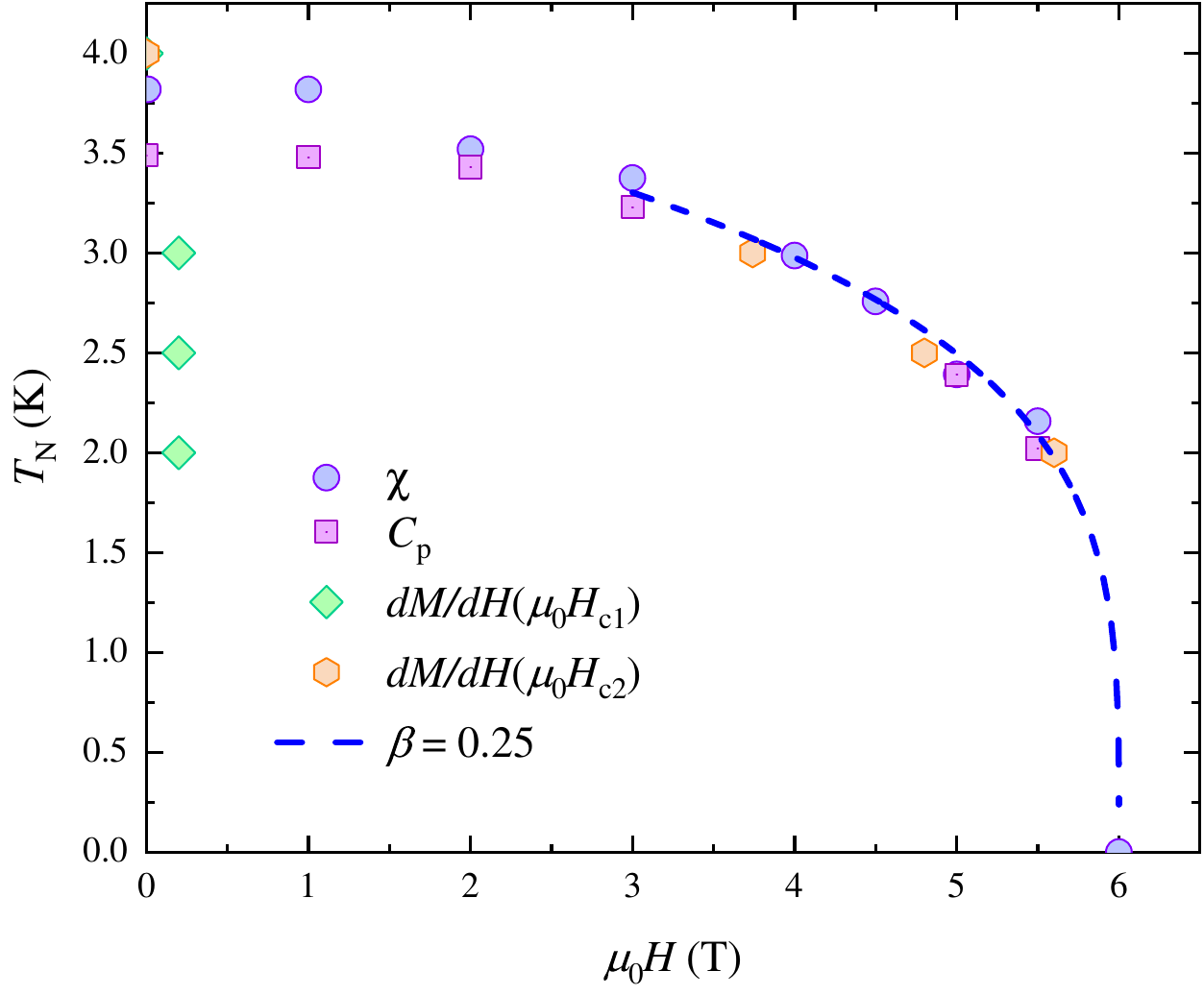}
	\caption{Phase diagram of
		FeP$_{3}$SiO$_{11}$ with critical points estimated by
		various methods as indicated by different legends. The dashed line shows the critical behavior of $T_{N}$ with a critical exponent of $\beta$ = 0.255(21).	}{\label{FPSOph}}.
\end{figure} 
\begin{figure*}
	\centering
	\includegraphics[width=\textwidth]{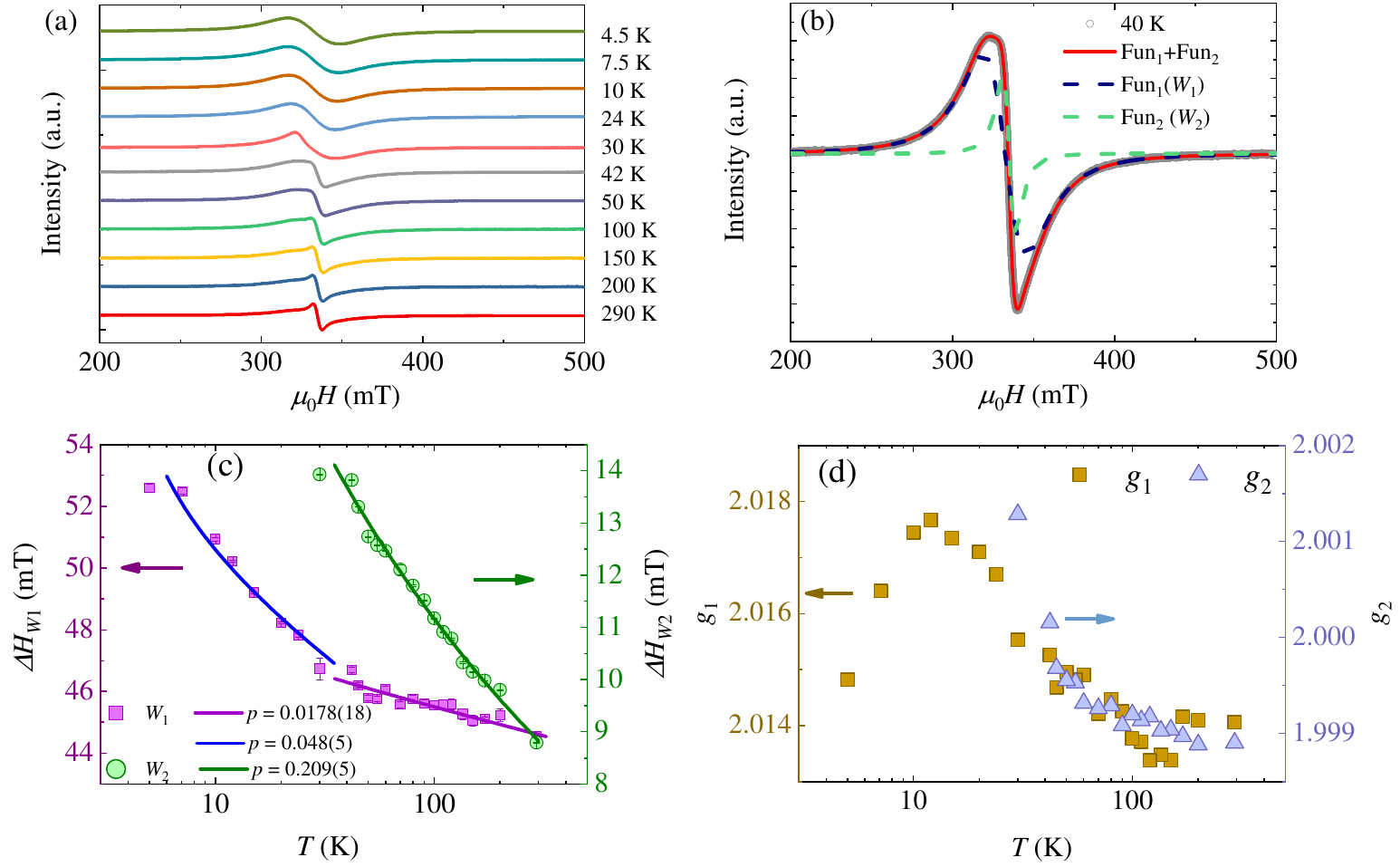}
	\caption{(a) Derivative of the ESR absorption spectra of FeP$_{3}$SiO$_{11}$ at selected temperatures. (b) Derivative of an ESR absorption spectrum at 40 K that is modeled by the combination of two Lorentzian functions, depicted by the red line. Each constituent Lorentzian function is illustrated by the blue and green dashed lines, corresponding to broad and narrow lines, respectively. (c) 
		Temperature dependence of ESR linewidths  is shown for broad (left y-axis) and narrow (right y-axis) lines, respectively. The solid lines
		indicate critical divergences  of linewidth for $T<T_{N}$ according to $\Delta H \propto ((T-T_{N})/T_{N})^{-p}$.   (d) 
		Temperature dependence of $g$-factor for two components denoted as $g_{1}$  and $g_{2}$  on the left and right $y$-axes, respectively.	}{\label{FPSO4}}.
\end{figure*} 
Magnetic entropy was calculated by integrating $C_{\rm mag}/T$ with respect to temperature. To capture the entropy release below 2 K, $C_\mathrm{mag}$ was extrapolated down to zero temperature via the observed low-$T$ power-law. The result is shown in the inset of Fig.~\ref{FPSO3}(b). Total magnetic entropy is found to be 12.67 J/mol$\cdot$K at 30 K, which is  15 \text{\%}  lower than the expected value $R$ $\ln(6)$ = 14.89 J/mol$\cdot$K for $S$ = 5/2 spins. The missing entropy is attributed to either the presence of short-range magnetic correlations persisting well above $T_{N}$ or a potential overestimation of the lattice contribution due to the limitations of the model in Eq.~\ref{eqn:debye} \cite{Khatua2021,PhysRevB.107.214411}.  Moreover, it is noted that below $T_{N}$, the calculated entropy represents 72\text{\%} of the total entropy, indicating that the release of the remaining entropy (approximately 28\text{\%}) occurs above $T_{N}$. However, the lack of a few percentage of entropy because of short-range order persisting above the transition temperature is typical for frustrated magnets \cite{PhysRevB.105.094439,Khatua2021}. This observation is corroborated by ESR results presented in  subsection \ref{subsec:esr}.\\
 In order to explore the effect of magnetic field on the
 antiferromagnetically ordered ground state of FPSO, we also performed specific heat measurements in several magnetic
 fields as shown in Fig.~\ref{FPSO3}(c). Consistent with magnetic susceptibility results (Fig.~\ref{FPSO2}), we observed
 that with increasing magnetic field, the transition temperature shifts toward lower temperatures. Surprisingly,
 the long-range antiferromagnetic order is completely suppressed in magnetic fields $\mu_{0}H$ $\leq$ 6 T which is close to the critical field $\mu_{0}H_{\rm c2}$ estimated from $dM/dH$ [Fig.~\ref{FPSO2}(d)].
A similar shift of transition temperature with magnetic field has also been observed in specific heat studies of several frustrated materials \cite{PhysRevB.104.094428,Lin2021,PhysRevLett.119.037201}.
To obtain a more detailed understanding of the magnetic ground state above $\mu_{0}H_{\rm c2}$, magnetic specific heat was measured for $\mu_{0}H\geq 6$ T and  is shown in Fig.~\ref{FPSO3}(d).
Interestingly, the magnetic specific heat shows a broad maximum with a double-peak structure around 3 K in a magnetic field of 6 T. Upon increasing the magnetic field, this broad maximum gradually shifts towards higher temperatures. The inset in Fig.~\ref{FPSO3}(d) shows that the position of the broad maxima increases linearly with the magnetic field. In a fully polarized state, gapped magnon excitations are created by flipping spins in the forced ferromagnetic background. In a nearly-polarized state, the confluence of magnetic anisotropies and short-range correlations may result in a small admixture of a disordered state, i.e., effectively free spins.  
In order to confirm the origin of a broad maximum in the disordered part of $C_{\rm mag}$, we attempt to analyze the temperature dependence of $C_{\rm mag}$ using a two-level Schottky model, taking into account the presence of free spins  [Fig.~\ref{FPSO3}(d)] \cite{PhysRevB.72.104501}. It has been noted that the broad maximum cannot be adequately explained by a simple Schottky model of free spins, instead points towards a more intricate correlated picture. This indicates the presence of substantial exchange correlations between $S$ = 5/2 spins of Fe$^{3+}$ ions in high magnetic fields. However, such broad maxima can be attributed to a field-induced disordered state as observed in spin liquid candidates Na$_{2}$Co$_{2}$TeO$_{6}$ \cite{Lin2021}  and K$_{2}$Ni$_{2}$(SO$_{4}$)$_{3}$ \cite{PhysRevLett.127.157204}.\\  Figure.~\ref{FPSOph} shows the temperature versus magnetic field phase diagram obtained via  magnetization and specific-heat measurements. This suggests that the presence of a zero-field spin configuration with antiferomagnetic ground state below $\mu_{0}H_{\rm c1}$. However, with increasing field, spin reorientation may lead to another field-induced antifferomagnetic phase that persists up to 6 T. Near the critical field of $\mu_{0}H_{\rm c2}$, the transition temperature $T_{N}$ exhibits an order-parameter form $T_{\rm N}(H)$ = $T_{\rm N}(0)$ (1-$H/H_{c}$)$^{\beta}$ with $\beta = 0.255(21)$ and $\mu_{0}H_{\rm c}$ = 6 T. Peculiarly, the obtained value of $\beta$ is close to the expected for the 2D
	XY model, indicating effectively 2D critical behavior of correlated $S$ = 5/2 spins of Fe$^{3+}$ ions in FPSO near this transition \cite{STBramwell1993,PhysRevB.107.064409} .
  
 \subsection{Electron spin resonance}\label{subsec:esr}
 In order to gain further insight into electron spin correlations in FPSO,
 ESR measurements
 were performed  in the X-band range at a fixed frequency of 9.4 GHz down to 4.5 K. Figure~\ref{FPSO4}(a) depicts the derivative of the ESR spectra at a few representative temperatures.
At $T$ $\geq$ 30 K, the ESR spectra can be accurately modeled as a sum of two Lorentzian lines, one broad ($W_1$) and one narrow ($W_2$), as shown in Fig.~\ref{FPSO4}(b). 
 The presence of two components in the absorption spectra of single-phase powder samples would suggest either two different Fe$^{3+}$ ion sites or distinct magnetic correlations in the in-plane and out-of-plane directions of the honeycomb lattice. As our Rietveld refinement revealed the presence of just a single Fe$^{3+}$ crystallographic site (Table~\ref{table}, Fig.~\ref{FPSO1}) we can eliminate the first possibility of multiple Fe$^{3+}$ ion sites. The two components should thus be attributed to distinct magnetic correlations in the in-plane and out-of-plane directions of the honeycomb lattice, i.e., to spin-fluctuation anisotropy \cite{PhysRevB.103.184413,PhysRevB.90.064419}, similar to that observed in  the kagome lattice antiferromagnet CsMn$_{3}$F$_{6}$(SeO$_{3}$)$_{2}$ ($S$ = 2). In this regard, future studies on single crystals could reveal valuable further insights.\\ The estimated ESR linewidths corresponding to the broad ($\Delta H_{W_{1}}$) and narrow ($\Delta H_{W_{2}}$) lines are shown in Fig.~\ref{FPSO4}(c), where the left $y$-axis represents $\Delta H_{W_{1}}$ and the right $y$-axis depicts $\Delta H _{W_{2}}$.  Above 30 K, the linewidth of the $W_{1}$ line exhibits only a weak temperature dependence, whereas the linewidth of the $W_{2}$ line  increases more quickly. Such a temperature dependence of linewidth  above the Weiss temperature indicates the existence of short-range spin correlations, a common feature observed in frustrated magnets \cite{PhysRevLett.127.157202,khatua2024magnetism}. It is noteworthy that below 120 K ($\gg$ $\theta_{\rm CW}$), the CW fit shows a deviation from the measured $\chi(T)$ which correlates with the observed temperature dependence of the ESR linewidth. \\
  Using the $T_{N}$ = 3.5 K obtained from thermodynamic data, we observed that the estimated ESR line width follows a $\Delta H_{W_{1}}$/$\Delta H_{W_{2}}$ $\propto$ ($T$/$T_{\rm N}-1$)$^{-p}$ critical-like behavior with $p$ = 0.0178(18) and $p$ = 0.209(5) corresponding to the $W_{1}$ and $W_{2}$ lines, respectively.
 In transition metal-based systems, the temperature dependence of the ESR linewidth is associated with spin--spin correlations, often resulting in a non-zero exponent value, defined as $p$ \cite{PhysRevB.80.054406}. At temperatures $T>|\theta_{\rm CW}|$, a finite value of $p$, albeit very small, suggests the presence of short-range spin correlations similar to those observed in a classical spin liquid candidate  Li$_{9}$Fe$_{3}$(P$_{2}$O$_{7}$)$_{3}$(PO$_{4}$)$_{2}$ \cite{PhysRevLett.127.157202}. It is noteworthy that, as the temperature is reduced below $T\leq$ 30 K, the narrow line vanishes, leaving only the broad line, which persists down to the lowest measured temperature. The rapid increase in the corresponding linewidth of the $W_{1}$ line below 30 K indicates a slowing down of spin dynamics, likely due to the onset of long-range magnetic order at $T_{N}$.
 In the temperature range 4.5 K $\leq$ $T$ $\leq$ 30 K, the linewidth for the $W_{1}$ line shows critical behavior with a different exponent of $p = 0.048(5)$.  Such a divergence of the  linewidth at lower $T$ is in line with short-range spin correlations, which also reflect the origin of the missing a few percentage of  the total entropy in thermodynamic experiments. \\
 The estimated temperature dependence of $g$-factors $g_{1}$ (left $y$-axis) and $g_{2}$ (right $y$-axis) for the $W_{1}$ and $W_{2}$ ESR lines are depicted in Fig.~\ref{FPSO4}(d).   Above  $T$$ \gg$ 160 K, both are approximately constant and equal to $g\approx$ 2 typical for Fe$^{3+}$ ions
 with a 3$d^{5}$ ( $S$ = 5/2) electronic configuration  in octahedral O$^{2-}$
 coordination. This indicates the temperature range 150 K $\leq$ $T$ $\leq$ 300 K is a paramagnetic regime, consistent with $\chi({T})$ data. Upon lowering the temperature below 150 K, $g$ factors of both lines start to increase, implying the presence of local spin correlations above $|\theta_{\rm CW}|$. This behavior is frequently observed in frustrated magnets, where frustration and geometric constraints bring about enhanced spin fluctuations \cite{PhysRevB.105.094439}. 
 Notably, a decrease in the g-factor has been noted below 10 K for the $W_{1}$ lines, which could be associated to the emergence of distinct spin correlations attributed to the proximity to
  long-range magnetic order. Similar to FPSO, the realization of multi-stage spin correlations above $T_{N}$ has been observed in various frustrated magnets \cite{PhysRevB.103.214447,PhysRevB.105.094439}. These multi-stage correlations indicate that the system may develop spin correlations of  spatially distinct character.
 \vspace*{-0.3 cm}
   \section{Conclusion}
 In summary, we have successfully synthesized and investigated the crystal structure and ground state properties of the previously unexplored transition metal-based compound FeP$_{3}$SiO$_{11}$.
 Using magnetization, specific heat, and ESR measurements, we determined that it crystallizes in a trigonal crystal structure with the space group R$\bar{3}$c, wherein Fe$^{3+}$ ions with $S$ = 5/2 spin form a nearly-perfect 2D honeycomb lattice in  crystallographic $ab$-plane, with no site disorder between constituent atoms.
 Our thermodynamic data reveal that the antiferromagnetically-coupled 
 $S$ = 5/2 spins host a  long-range magnetic order state below  $T_{N}$ = 3.5 K in zero-field. The competition of intra-planar nearest-neighbor and  inter-planar further-neighbor exchange interactions stabilizes a long-range magnetic ordered state in FPSO, as confirmed by our DFT calculations.    As the magnetic field increases, $T_{N}$ shifts towards lower temperatures, and its suppression continues gradually up to a critical field of approximately $\mu_{0}H_{\rm c}$ = 6 T.  The correlation between the strength of the applied magnetic field and $T_{N}$ implies the presence of partially-ordered magnetic moments below $T_{N}$.
 Similar to  magnetization and specific heat data, the presence of a critical field around $\mu_{0}H_{c}$ is revealed by a cusp-like feature observed in the field-derivative of magnetization ($dM/dH$) below $T_{N}$.  Furthermore, an additional cusp appears at $\mu_{0}H_{\rm c1}$ in $dM/dH$ below $T_{N}$, associated with a spin-flop-like transition. Interestingly, the appearance of a broad maximum at $\mu_{0}H_{c}$ suggests a possible scenario of short-range spin-correlations due to a field-induced disordered state on top of a forced-polarized state. ESR results provide a preliminary indication of in-plane and out-of-plane magnetic-correlation anisotropy, which might play a crucial role in stabilizing the magnetic long-range magnetic ordered state in FPSO. Furthermore, the critical divergence of the ESR linewidth suggests the presence of short-range spin correlations well above $T_{N}$, which is consistent with the observed small percentage of missing magnetic entropy compared to the expected value. This observation is also consistent with competition between intra-planar nearest-neighbor and the next-nearest-neighbor inter-planar magnetic interactions, as supported by the extracted moderate frustration parameter, $f \approx 4$. 
 Below $T_{N}$,  magnetic specific heat shows a power-law behavior, setting FPSO apart from other 3D antiferromagnets. Indeed, the temperature versus magnetic field phase diagram indicates that near the critical field $\mu_{0}H_{c2}$, the critical exponent $\beta$, which corresponds to the temperature dependence of the order parameter, approaches $\beta = 0.255(21)$. This value is close to that expected for a 2D XY magnets, implying peculiar 2D critical behavior of correlated spins in FeP$_{3}$SiO$_{11}$ despite its being a manifestly 3D magnetic system. \vspace*{0.3 cm}
   \section*{Acknowledgments}
     P.K. acknowledges the funding by the Science and
    Engineering Research Board, and Department of Science and Technology, India through Research Grants. The work at SKKU was supported by the National Research Foundation
    (NRF) of Korea (Grant no. RS-2023-00209121, 2020R1A5A1016518). Computing resources were provided by the STFC Scientific Computing Department’s SCARF cluster. 
    M.G. acknowledges the financial support of the Slovenian Research and Innovation Agency through Program No. P1-0125 and Projects No. Z1-1852, N1-0148, J1-2461, J1-50008, J1-50012, and N1-0356. 
\bibliography{FPSO}
\end{document}